# The XPS peak structure of condensed aromatic anhydrides and imides

M. Jung[1*], U. Baston[1], T. Porwol[2], H.-J. Freund[2], and E. Umbach[3]

[1] *4. Physikalisches Institut, Universität Stuttgart, D-70550 Stuttgart, Germany*
[2] *Fritz-Haber-Institut der Max-Planck-Gesellschaft, D-14195 Berlin, Germany*
[3] *Experimentelle Physik II, Universität Würzburg, D-97074 Würzburg, Germany*

**Abstract**

Photoelectron spectra of aromatic anhydrides and imides like PTCDA (perylene-tetracarboxylic dianhydride), PTCDI (perylene-tetracarboxylic diimide), and NDCA (naphthalene-dicarboxylic anhydride) on smooth single crystal surfaces show complex double peak structures in both the $O_{1s}$ and the anhydride $C_{1s}$ range. The peak intensity ratios cannot be simply explained by different chemical states of the atoms but are strongly influenced by intense shake-up satellites. Semiempirical SDCI calculations can quantitatively reproduce the experimental spectra of multilayers, provided that not only intramolecular but also intermolecular shake-up excitations are taken into account. These calculations give further insight into the process of electronic charge transfers in large organic molecules and molecular films. A variation of the molecular structure by reducing the size of the aromatic part (NDCA instead of PTCDA) or by using different functional groups (PTCDI instead of PTCDA) leads to a reduced donor/acceptor strength and hence to drastically reduced satellite intensities.

* E-mail address: michael.jung@gmx.de

# 1. Introduction

In the last years the adsorption of large organic molecules and polymers on metal and semiconductor surfaces has received considerable attention *[1-10]*. Moreover, there is growing interest in technological applications of these organic/inorganic hybrid systems *[11-13]*, for example as dielectrical and insulating protective coatings *[14-16]*, sensors *[17,18]*, and organic light emitting diodes *[19-24]*. Therefore, the electronic and geometric structure of the interface, e.g. the chemical bonding of the molecular adsorbate, and about the geometric and electronic structure is very important. Surface sensitive methods such as photoelectron spectroscopies (PES) have been useful tools for the investigation of these problems.

Much experimental and theoretical work has been performed on polymers and on model molecules for polymers, i.e. small oligomers or monomers with suitable endgroups. In particular organic anhydrides and imides are interesting model systems. From the experimental point of view monomers are easier to purify and can be adsorbed in a very controlled way. Therefore the results are less ambiguous and easier to interpret. On the theoretical side calculations can be performed on an higher sophisticated level so that experimental results and calculations can be compared directly *[25-28]*. The experimental XP spectra of such systems show similar satellite structures as observed earlier for small inorganic adsorbates *[29-31]*. These satellites cannot be interpreted as initial state effects because of a modified chemical environment but are due to strong dynamic multi-electron shake-up excitations in the PES final state *[32]*. Significant work based on different theoretical models has been done to describe this process in detail *[33-39]*. Moreover, the satellite peaks observed for polar and nonpolar organic molecules have also been quantitatively explained by shake-up intensities accompanying ionization of the system *[40]*.

In this paper we will discuss and quantitatively explain the influence of dynamic final state effects on the XP spectra of organic anhydrides and imides. Three molecules - PTCDA, PTCDI and NDCA (see Fig. 1) - are choosen to study in detail the influence of the aromatic system and of the functional groups. Thin films (about 20 ML) of these molecules has been prepared by temperature controlled vacuum sublimation on Ni(111) and Ag(111).

PTCDA (perylene-tetracarboxylic dianhydride) consists of an aromatic perylene core and two anhydride groups. The single molecule shows $D_{2h}$ symmetry which is slightly disturbed upon crystallization. To determine the influence of the functional groups on the shake-up satellites we further examined PTCDI (perylene-tetracarboxylic diimide). PTCDI consists of imide groups instead of anhydride groups and is isoelectronic to PTCDA (also with $D_{2h}$ symmetry). As the size of the aromatic core remains the same, changes in satellite intensities should primarily be influenced by the different functional groups. NDCA (naphthalene-dicarboxylic anhydride) is half a PTCDA: it consists of a naphthalene aromatic core with a single anhydride group. The molecular symmetry is therefore reduced to $C_{2v}$. The influence of the reduced size of the aromatic system should be seen in the spectra of adsorbed multilayers and in the according calculations.

## 2.  Experimental and computational details

The measurements were performed in a UHV system (VG ESCALAB MK II) consisting of separate preparation and analysis chambers which are equipped with temperature controlled evaporation cells for sample preparation, a quadrupole mass spectrometer (Balzers QMG 511), a three-grid LEED optics, a twin-anode X-ray source, a He discharge lamp, and a 150° spherical analyser for photoelectron spectroscopy. The sample temperature could be adjusted in the range from 100 K to 1200 K. All experiments were made at a base pressure of $1*10^{-10}$ mbar or below. The substrates Ni(111) and Ag(111) were cleaned by repeated argon etching and annealing cycles until no XPS impurity signals were detected and LEED showed a perfect substrate pattern. Afterwards the organic adsorbate was deposited by vacuum sublimation; both, adsorption rate and dose were controlled by the quadrupole mass spectrometer and calibrated using XPS and thermal desorption (TDS). After transfering the prepared sample into the analyser chamber XPS measurements were performed. For the Ni substrate the Mg-$K_\alpha$ source could not be used in the $O_{1s}$ energy range because of overlapping Ni Auger transitions. Therefore, in this case all XPS measurements were made using Al-$K_\alpha$ radiation in spite of the larger linewidth. It is remarkable that no detectable X-ray-induced radiation damage occurred within several hours for all molecular systems studied.

The quantum chemical calculations were performed using the semiempirical all-valence-electron closed shell formalism. Starting with crystal and force-field data for the molecular parameters a geometric optimization of the plane molecules has been done with the MOPAC program package *[41]* using MNDO, AM1 and PM3 algorithms. This optimization was necessary as UPS *[42,43]*, LEED *[44-46]* and STM *[44-46]* experiments suggest that the crystal data *[47,48]* has to be slightly adapted for condensed layers. The geometries obtained within the PM3 calculations are in best agreement with the spectroscopic results. Based on the optimized geometries CNDO/S-CI and INDO/S-CI calculations *[49-54]* including up to doubly excited states and using the Nishimoto-Mataga formula of the two-center repulsion integrals *[55,56]* have been carried out. Excited state computations were performed with the „equivalent core" approximation for closed shell species assuming only singlet coupling between all valence electrons. The CI of the core-ion state included the 700 lowest energy excitations selected out of 5000 configurations of proper symmetry for doubly excited states. Relative intensities were obtained within the „sudden approximation" by projection of the „relaxed" ionic state wavefunctions onto those of the „frozen" ionic state given by the one-electron orbitals of the neutral system *[57-59]*. In order to compare shake-up intensities by both intra- and intermolecular charge transfer, calculations were performed for single molecules as well as for sandwich clusters of three molecules.

## 3. Results and discussion

### 3.1 Condensed layers at RT

Fig. 2 shows $C_{1s}$ and $O_{1s}$ spectra of multilayers (about 20 ML) of PTCDI (top), PTCDA (middle) and NDCA (bottom) on Ni(111) prepared at room temperature. The spectra are normalized to the same peak height. The $C_{1s}$ spectra in Fig. 2a consist of two well separated peak structures with peak maxima at about 285.3 eV for peak 1 and 288.3 - 289.3 eV for feature 2 (hatched). Perylene spectra as well as known $C_{1s}$ binding energies of other organic compounds *[60,61]* let us identify peak 1 as resulting from carbon atoms of the aromatic part and peak 2 from carbon atoms of the anhydride and imide endgroups, respectively.

It is conspicuous that there is hardly an energy shift (<0.1eV) of the aromatic peak 1 between PTCDI, PTCDA and NDCA. We cannot separate the contributions of the individual aromatic carbon sites $C_1$, ... , $C_{10}$ (see Fig. 1) within the experimental resolution, but the weak asymmetry at the low binding energy side - hardly seen in Fig. 2 - probably indicates the slightly different chemical environment of the various carbon atoms in the aromatic part of the molecules.

The well separated feature 2 of the $C_{1s}$ spectra can be attributed to the carbon atoms $C_{11}$ and $C_{12}$ of the anhydride (PTCDA, NDCA) and imide (PTCDI) groups, respectively. Both, the binding energy and the line shape of the anhydride/imide structures vary significantly for the three molecules. From NDCA over PTCDA to PTCDI the peak maximum shifts by more than 1 eV to lower binding energies. Moreover, the anhydride peak consists of more than one component although both carbon atoms $C_{11}$ and $C_{12}$ are chemically identical. The shoulder on the high binding energy side is significantly more intense for PTCDA than for NDCA or PTCDI, but it can be clearly seen in the $C_{1s}$ spectra of all three molecules. It is noted that the stoichiometric ratio of the carbon atoms in the aromatic part to those in the functional groups (which is 5:1) is only reproduced by the corresponding peak intensities (areas) when the whole hatched structure 2 (peak plus shoulder) is assigned to the carbon atoms of the functional groups. This clearly indicates that peak and shoulder belong together. As stated below the shoulders are satellites of the main anhydride or imide peak 2.

Additional information comes from the corresponding $O_{1s}$ spectra of PTCDI, PTCDA and NDCA shown in Fig. 2b. For all three molecules the spectra exhibit double peak structures with an energy separation of about 1.9 eV. The intensity of the high-energy component decreases form PTCDA over NDCA to PTCDI. While this component has considerable intensity in the case of PTCDA and NDCA, it is only a small shoulder in the PTCDI spectrum. If we interpret the low-energy peak 1 as usual *[60,61]*, namely resulting from the carbonylic oxygen atoms, and peak 2 at higher binding energy as derived from the central oxygen atoms, their stoichiometric intensity ratio should be 2:1 for PTCDA and NDCA, and 2:0 for PTCDI. It is obvious that neither the experimental intensity ratio which is close to 5:4 (PTCDA) and 3:2 (NDCA), respectively, nor the PTCDI structure 2 at all correspond to this simple stoichiometric picture.

The explanation is that peak 2 rather consists of a superposition of several components which are only slightly shifted in energy. The increased peak width of peak 2 compared with peak 1 gives additional evidence for this interpretation. One of these components is clearly assigned to the central oxygen atoms, at least in the case of PTCDA and NDCA. The nature of the other is based either on a chemically modified initial state of a fraction of the oxygen atoms or on dynamic multi-electron effects in the photoemission final state (shake-up satellites).

An electronic asymmetry of the functional groups induced by chemical bonding, as proposed for the interface of small adsorbed anhydrides and imides on metal surfaces *[62-64]*, can be excluded in the present case, i.e. for multilayers, because of several reasons:

It is very unlikely that in condensed layers, which interact mainly by Van-derWaals forces, the carbonylic oxygen atoms experience such a strong chemical interaction that is needed for the large experimental energy shifts observed. Moreover, in the case of strong interaction of the carbonyl atoms (where oxygen would act as an electron donor) a corresponding energy shift of opposite sign would be expected in the $C_{1s}$ spectrum, which is not found. Finally, other spectroscopic results from TDS and NEXAFS clearly indicate that the intermolecular bonding is weak *[43,65,66]*.

Thus we come to the conclusion that the above observations result from dynamic multielectron („shake-up") effects in the photoemission final state. In other words, the shoulder in the anhydride $C_{1s}$ feature 2 and the additional components in the $O_{1s}$ spectrum around 533-536eV belong to relatively intense shake-up satellites whose intensities depend on the ionized atom. As shown in detail below in section 3.2.2 and as seen in Fig. 2 mainly the carbon atoms of the imide/anhydride groups and the carbonylic oxygen atoms exhibit significant satellite structures. This also explains the non-stoichiometric peak ratio in Fig. 2b as a loss of intensity in peak 1 and a gain of intensity in peak 2 by superposition of satellite contributions and intensity from the central anhydride oxygen atom.

In PTCDI this central oxygen atom is replaced by a NH group; thus the $O_{1s}$ spectrum only represents the contribution of the carbonyl oxygen atoms (Fig. 2b) and hence feature 2 can unambiguously be assigned to their shake-up satellites alone.

The XP spectra of multilayers of the investigated aromatic molecules do not depend on the selected substrate. They are identical for adsorption on the smooth (111)-surfaces of nickel and silver and even on the more corrugated surface of [7x7] reconstructed Si(111) *[67]*.

## 3.2  Results of the calculations

### 3.2.1  Molecular orbitals

Fig. 3, 4 and 5 present the local distributions of the upper five occupied and the lower four unoccupied molecular orbitals (MOs) of PTCDI, PTCDA and NDCA together with their numbers and symmetries from LCAO-HF INDO/S calculations within the PM3-optimized geometry. The sizes of the circles indicate the contributions from the corresponding atomical orbital coefficients to each MO. The phase (+/-) of the wavefunctions is distinguished by open and filled circles, respectively. In each figure, the left column describes the ground state of the neutral molecule. The molecular symmetry is $D_{2h}$ in the cases of PTCDI and PTCDA, but $C_{2v}$ for NDCA. In the middle and right columns corresponding data for the $C_{11}$ and $O_{14}$ core ionized molecules are given. Several conclusions can be drawn immediately:

1) The sequence of MOs is identical for PTCDA and the isoelectronic PTCDI in the considered energy range. The LCAO coefficients and the resulting local electron distributions within the molecules are (almost) the same. This finding is valid for the neutral as well as for the ionized molecules. Apparently, the replacement of the anhydride oxygen atom by the imide group has only a small influence on the ground state wavefunctions of the free molecule.

2) As expected for aromatic molecules the highest occupied and lowest unoccupied MOs are of π-type. The occupied orbitals of the neutral molecule are predominantly located on the aromatic part, i.e. the contributions of the functional group atoms to these orbitals are only small. For chemisorbed molecules we note that especially the MOs from HOMO to HOMO-2 of PTCDA/PTCDI and HOMO to HOMO-1 of NDCA are strongly modified by the chemical bonds to the substrate *[42,43]*. The unoccupied orbitals are slightly more delocalized and have increased contributions from the functional groups, since particularly the LCAO $C_{11}$ and $C_{12}$ coefficients are more significant. Thus one may expect that the coupling to a metal substrate leads to an electron transfer to the (energetically lowered) first unoccupied MOs with participation of the whole molecule (both aromatic part and functional groups). Differences in the spectral features between a (sub)monolayer on metals with a high d-electron density at the Fermi level (e.g. Ni) and on metals with a deep d-band and small s-electron density at $E_F$ (e.g. Ag) may result from this. A detailed discussion including UPS and IPS data will be given elesewhere *[68]*.

3) The MOs of the ionized molecule determine the photoemission final state. The core ionized atoms in the middle and right columns of Fig. 3-5 are $C_{11}$ and $O_{14}$ of the functional groups (see Fig. 1). Apart from the induced symmetry reduction from $D_{2h}$ and $C_{2v}$, respectively, to $C_s$, the local distributions of the MOs also differ. While the LCAO contributions of the functional group atoms are only slightly changed for the HOMO, they increase drastically for the LUMO and LUMO+1 in all three molecules. Therefore low-energy shake-up excitations involving these orbitals should influence the charge distribution within the molecule, as discussed in the next section.

### 3.2.2 SDCI calculations

The results of INDO/S-CI calculations discussed in the following allow a more quantitative insight into the XPS satellite structure. Analogous to the ground state calculations of the previous section the CI calculations have been done mainly using the PM3-optimized geometry of the neutral molecule. Of course, if only a single molecule is considered the result is restricted to intramolecular excitations. Especially in multilayers this assumption may be too restrictive. In order to allow additional intermolecular processes a symmetrical layered pile of three molecules was also examined, using an intermolecular distance of 3.22 Å, which is the same value as in PTCDA crystals *[47,48]*. This approach is necessary to get a sufficiently good agreement with the experimental data as discussed below.

Tab. 1 shows the energy eigenvalues and intensities for those first single-molecule CI states which mostly contribute to the total intensity of core ionized PTCDA. The ionized atoms in the table are $C_1$, $C_3$, $C_{11}$, $O_{13}$ and $O_{14}$. They are denoted according to Fig. 1: $C_1$ and $C_3$ are aromatic atoms and $C_{11}$, $O_{13}$ and $O_{14}$ represent the anhydride group. The first columns of the table give the numbers of the CI states, their energy shifts relative to the unrelaxed ground state and their percentage relative to the total intensity; the coefficients of the dominating single and double excitations to the appropriate wavefunctions are given in the last column. The CI state #1 always denotes the „main peak" which essentially corresponds to the reorganized ground state configuration, which is given as () in the table. The energy of state #1 is the correlation energy shift of the ionized system.

In the case of carbon ionization only the aromatic atoms $C_1$ and $C_3$ and the anhydride atom $C_{11}$ (and of course their symmetry-eqivalent analogues $C_6$, $C_4$, $C_{12}$) carry high shake-up probabilities leading to distinct satellite peaks. All other core atoms ($C_2$, $C_7$, $C_9$, $C_{10}$ and their symmetric equivalents) lead only to a small, unstructured satellite spectrum of low intensity and are therefore not explicitly given in the table. In all cases, the first excited CI state #2 carries the highest intensity which varies from about 7% for $C_1$ up to 34% for the anhydride carbon atom $C_{11}$. The energy splitting between this satellite and the CI ground state #1 is about 2.5 eV and 1.5 eV, respectively, large enough that the two states should be distinguishable in the experimental spectra.

Fig. 6 shows the composed CI calculated $C_{1s}$ spectrum of PTCDA (Fig. 6b,c) and the corresponding experimental XP multilayers spectrum (Fig. 6a) taken with Al-$K_\alpha$ radiation. The theoretical spectrum represents the sum of a set of Voigt functions (FWHM: Lorenz=0.5 eV, Gauß=0.65 eV) for the aromatic and the anhydride carbon atoms (marked 1 and 2, respectively). Each set consists of the main peak and of satellites with a higher intensity than 0.0001 out of the first 700 satellites of low energy according to the CI calculation. The relative areas of the two sets are given by the stoichiometric ratio of the carbon atoms in the PTCDA molecule as $C_{1+2+...+10}$:$C_{11+12}$ = 5:1. As the experiment cannot resolve the binding energies of the different aromatic carbon atoms, one single binding energy was used for all atoms of the aromatic core. This approximation is apparently not quite correct as can be derived from the large width and slight asymmetry of the experimental main peak, which cannot exactly be reproduced by a single Voigt function.

The experimental energy splitting between the aromatic peak 1 and the anhydride peak 2 is 3.55 eV. By consideration of the slight differences in the correlation shifts between aromatic and anhydride carbon atoms this is the same value as calculated within the relaxation-potential model *[69]* for the ionized system, originally derived by Siegbahn et al. *[70,71]* for the ground state.

The first shake-up satellites of $C_1$ and $C_3$ (marked as structure $1^*$ at about 2.5 eV above the aromatic main peak 1) are not well separated at the low-energy side of the anhydride peak 2 in the original experimental spectrum 6a, but can be clearly seen in a deconvoluted spectrum. The other aromatic satellites are superimposed to the anhydride spectrum and therefore not distinguishable. The anhydride carbon $C_{11}$ leads to a much higher shake-up intensity in both, experiment and calculation. Whereas the energy splitting between the anhydride main peak 2 and its first, most intense satellite $2^*$ is well reproduced by the shake-up calculation of a single molecule (Fig. 6c), this approach apparently overestimates the satellite intensity. However, we must consider that the spectrum also contains a significant contribution of aromatic satellites in the energy region of peaks 2 and $2^*$. Small variations in this aromatic „background" by additional **inter**molecular excitation processes, e.g. in the more extended electron system of a molecular cluster, will affect the intensity ratio of peak 2 and $2^*$, although the major anhydride satellite structure remains determined only by **intra**molecular excitations. The resulting spectrum of such a configuration of three molecules arranged in a symmetric pile is shown in Fig. 6b. Due to the smaller energy distances of the orbitals of this cluster the aromatic shake-up structure is slightly different to that of the single molecule and hence changes the background around the anhydride peaks yielding a better agreement with the experiment.

The shake-up structures of the oxygen atoms also differ drastically: the ionization of the (central) anhydride atom $O_{13}$ does not yield intense low-energy shake-up satellites, whereas ionization of the carbonyl atom $O_{14}$ leads to significant satellites intensities. This different behaviour is not astonishing, if we consider the symmetries and LCAO coefficients of the MOs in the range around HOMO and LUMO as shown in Fig. 4. Especially the first shake-up excitation at about 2 eV above the main peak contains more than 17 % of the total intensity of the $O_{14}$ spectrum. The loss of intensity due to the entire satellite structure diminishes the carbonyl main peak by 30% and hence much more in comparison to the anhydride peak (12%). Furthermore, the binding energy of the first carbonyl satellite is in the same range as that of the anhydride main peak, which lies – according to the results of our calculations within the potential model of the completely relaxed ionized molecule – about 1.8 eV higher than the carbonyl peak. Both effects influence the carbonyl/anhydride peak ratio drastically such that an incorrect oxygen stoichiometry value may be derived.

Fig. 7 shows an $O_{1s}$ spectrum of PTCDA multilayers (dotted) together with three different simulations. In all cases the theoretical spectra consist of two Voigt-shaped components representing the carbonyl peak 1 and the anhydride peak 2. Both have identical peak parameters that are derived by fitting the low-energy side of the experimental peak 1. Below each spectrum the appropriate individual components are given. Following the argumentation above, the approach of a stoichiometrical carbonyl/anhydride intensity ratio of 2:1 without satellites cannot reproduce the experimental spectrum (Fig. 7a). The width and intensity of peak 2 are much too small. The calculated spectra of Fig. 7b stem from a CI calculation of a single molecule; the intense first carbonyl satellite at the position of the anhydride peak at 534 eV is clearly observed, but now the total intensity at peak 2 is overestimated. Furthermore the trailing edge of the experimental peak 2 is not well reproduced. If we perform a calculation based on a pile of three molecules (Fig. 7c) a much better agreement with the experimental data is achieved, as in the $C_{1s}$ case. Now the intense carbonyl satellites do not exactly coincide with the anhydride peak, but broaden the anhydride structure at both sides. Hence this enhanced calculation matches the experimental spectrum very well, both in intensity and line width.

Thus we must conclude that, although intramolecular excitations within a single molecule qualitatively describe the major features of the experimental XP spectra, the consideration of additional intermolecular effects appears to be necessary to reproduce the multilayers data quantitatively.

The dominating shake-up contributions to the wavefunctions of intense CI states are exclusively $\pi \rightarrow \pi^*$ transitions. It is worth noting that the participating occupied MOs are mainly located on the aromatic part (see Tab. 1, Fig. 4, and *[42]*; $\varphi_{70}$ labels the HOMO). In contrast to the neutral molecule the ionized molecules possess a large wavefunction amplitude of the LUMO and (LUMO+1) on the anhydride group. Especially the first, most intense satellites are dominated by HOMO→LUMO transitions for $C_{1s}$ and HOMO→LUMO as well as HOMO→(LUMO+1) transitions for $O_{1s}$ ionization, respectively. Therefore the shake-up process can be attributed to an intramolecular charge transfer since by this excitation electronic charge is moved from the aromatic part, acting as a donor, to the accepting anhydride group. This concept of a charge transfer shake-up has also been found for small aromatic and aliphatic organic molecules with and without oxygen containing functional groups *[40,72,73]*. The intensity of such charge transfer shake-ups is related to the donor/acceptor strength of the system *[74]*, which is influenced by two properties of the considered molecule: the size of the aromatic part and the nature of the functional group.

The differences in the experimental data of PTCDI, PTCDA and NDCA multilayers (see Fig. 2) can be interpreted exactly in this way and can be described by CI calculations on the single molecule as shown in Tab. 2 and 3: the CI states of PTCDI, PTCDA and NDCA are given after ionization of the carbon atom $C_{11}$ in Tab. 2 and of the carbonyl oxygen atom $O_{14}$ in Tab. 3 for comparison. The structure of these tables is the same as described above for Tab. 1, except for the normalization of the relative intensities and energies to those of the fully screened ionic state (= 0 eV, 100 %).

First we note that the replacement of the PTCDA anhydride oxygen atom by the isoelectronic imide group in PTCDI does neither alter the sequence of CI states nor the shake-up composition of their wavefunctions. This is not surprising as the involved MOs of PTCDA and PTCDI are very similar (see Fig. 3 and 4). The energy splitting between the CI states of PTCDA and PTCDI also remains fairly constant. Moreover, it is remarkable that the calculated intensitiy of the first PTCDI CI state #2 is clearly reduced with respect to that of PTCDA for $C_{11}$ as well as for $O_{14}$ ionization, in agreement with the experimental spectra of Fig. 2. This behaviour is related to the reduced donor/acceptor strength in the PTCDI molecule compared with PTCDA due to the different functional groups. The replacement of the anhydride group by the less polar imide group also leads to the smaller energy splitting between the aromatic and the anhydride PTCDI $C_{1s}$ peaks (see Fig. 2a) according to the potential model *[70,71]*.

The influence of the size of the aromatic part becomes obvious by comparing NDCA with PTCDA. The reduction of the aromatic system in NDCA also leads to a significant loss of satellite intensity but with a simultaneous increase of the energy splitting of the first CI states. A direct coordination of these NDCA CI states to those of PTCDA is, of course, not possible as the involved MOs of the neutral and ionized molecules are quite different in character and energy (see Fig. 4 and 5, as well as *[42,43]*). Significant additional contributions from other shake-up excitations occur for the NDCA CI state #2, e.g. from the (HOMO-1) to the LUMO. We emphasize that the intensity reduction of the experimental NDCA satellites (Fig. 2) is well reproduced by the calculated data given in the tables: the intensity of CI state #2 in the $C_{11}$ ionized NDCA molecule is hardly a third of that of PTCDA; the same satellite in the $O_{14}$ ionized state of NDCA has less than half the intensity compared to PTCDA. The explanation is obviously that the much smaller aromatic system of NDCA leads to a reduction of the donor capability and hence to the drastic spectral changes that can be observed.

## 4. Summary

$C_{1s}$ and $O_{1s}$ XPS data of condensed PTCDA, NDCA and PTCDI multilayers show complex peak structures, which cannot be understood in a simple initial state model of chemically different atoms. The spectra contain additional features which stem from multi-electron shake-up satellites as they are also well known for many other organic and inorganic molecules. A detailed analysis using SDCI calculations allows a quantitative explanation of the XPS structures. These calculations show shake-up excitations of significantly different intensities at the various atom sites of the molecule. Only the $C_{1s}$ peaks of the anhydride or imide carbon atoms and the $O_{1s}$ peaks of the carbonylic oxygen atoms are accompanied by intense satellites which can be attributed to specific multi-electron excitations. On the contrary, the ionization of the atoms of the aromatic part, of the anhydride oxygen atom and of the imide nitrogen atom leads only to weak unresolved shake-up structures.

Furthermore, a detailed consideration shows that the satellite intensity strongly depends on the size of the aromatic part and the polarity of the functional group. Both properties influence the molecular donor/acceptor strength which is important for the intramolecular excitation processes. The reduction of the satellite intensities from PTCDA to NDCA can be attributed to the smaller aromatic part, whereas the reduced polarity is the origin of the intensity reduction in the case of PTCDI.

For condensed layers a complete agreement between experiment and calculations is only achieved when intermolecular excitation channels are taken into account.

## 5. Acknowledgements

We like to thank Prof. N. Karl for providing the purified organic material and for valuable discussions. This work has been supported by the Deutsche Forschungsgemeinschaft through SFB 329. Two of us (E.U. and H.-J.F.) appreciate support by the Fond der Chemischen Industrie.

# Table captions

Table 1      Calculated energies and intensities for the first CI states of core ionized PTCDA. The wavefunctions are characterized by the configurations with the largest coefficients in the CI expansion. The HOMO is $\varphi_{70}$. For details see text.

Table 2      Comparison of the theoretical energy shifts and relative intensities of CI states for PTCDI, PTCDA and NDCA for core ionization of the carbonylic carbon atoms. The wavefunctions are characterized by the configurations with the largest coefficients in the CI expansion. The HOMO is $\varphi_{70}$ in the case of PTCDI and PTCDA, and $\varphi_{36}$ for NDCA. For details see text.

Table 3      Comparison of the theoretical energy shifts and relative intensities of CI states for PTCDI, PTCDA and NDCA for core ionization of the carbonylic oxygen atoms. The wavefunctions are characterized by the configurations with the largest coefficients in the CI expansion. The HOMO is $\varphi_{70}$ in the case of PTCDI and PTCDA, and $\varphi_{36}$ for NDCA. For details see text.

# Figure captions

**Figure 1**   The molecular structures of PTCDI, PTCDA and NDCA. Specific atom sites are labeled.

**Figure 2**   $C_{1s}$ spectra (a) and $O_{1s}$ spectra (b) of multilayers of PTCDI, PTCDA, and NDCA. The lines indicate the BE of the main peaks of PTCDA for both C1s and O1s; they are extended into the spectra of PTCDI and NDCA to emphasize relevant energy shifts. The anhydride and imide structures (2) in the $C_{1s}$ spectra are marked by hatching.

**Figure 3**   Number, symmetry assignment, and spatial distribution of the upper occupied and the lower unoccupied MOs of PTCDI from INDO/S calculations. The radii of the circles are proportional to the contributions (LCAO coefficients) of the corresponding atomic $p_z$ orbitals; open and full symbols indicate the phase of the wavefunctions.

**Figure 4**   Number, symmetry assignment, and spatial distribution of the upper occupied and the lower unoccupied MOs of PTCDA from INDO/S calculations. The radii of the circles are proportional to the contributions (LCAO coefficients) of the corresponding atomic $p_z$ orbitals; open and full symbols indicate the phase of the wavefunctions.

**Figure 5**   Number, symmetry assignment, and spatial distribution of the upper occupied and the lower unoccupied MOs of NDCA from INDO/S calculations. The radii of the circles are proportional to the contributions (LCAO coefficients) with $p_z$ character for π-orbitals and with a combination of $p_x, p_y$, and s character for σ-orbitals; open and full circles indicate the phase of the wavefunctions for π-orbitals only.

**Figure 6**   Comparison of an experimental $C_{1s}$ multilayers spectrum of PTCDA/Ni(111) (a) and the corresponding CI-calculated spectra of a symmetric pile of three

molecules (b) and one single molecule (c). The labels 1 and 2 denote the main peaks of the aromatic and anhydride part, the labels $1^*$ and $2^*$ denote the corresponding most intense shake-up satellites.

**Figure 7**  An experimental $O_{1s}$ multilayers spectrum of PTCDA/Ni(111) (dotted) is compared to a combination of two Voigt-shaped lines with the stoichiometrical intensity ratio 2:1 (a), and to the corresponding CI calculated spectra of one single molecule (b) and a symmetric pile of three molecules (c). The individual spectra of both different oxygen species are given as well as a sum spectrum. Label 1 denotes the carbonyl oxygen peak, label 2 the anhydride oxygen peak.

# Table 1:

| | PTCDA | | INDO-CI Calculation |
|---|---|---|---|
| # | Energy [eV] | Intensity [%] | Wavefunction |
| | | $C_1$ | |
| 1 | -0.94 | 71.4 | 0.95() - 0.11(69,69→75,75) |
| 2 | 1.53 | 6.8 | 0.89(70→71) + 0.14(67,70→71,71) + 0.13(70→72) + 0.12(70→73) + 0.12(70,70→71,72) |
| 10 | 3.67 | 4.4 | 0.71(64→71) - 0.24(68→72) + 0.22(70→77) - 0.20(68,70→71,71) + 0.16(70→76) |
| 13 | 4.16 | 1.4 | 0.58(70→77) - 0.29(64→71) + 0.18(69→73) - 0.15(69→72) + 0.15(70→72) - 0.13(68→72) |
| 17 | 4.43 | 1.4 | 0.48(69→74) - 0.17(70→77) - 0.15(69→73) + 0.14(70→76) - 0.14(67→72) - 0.13(64→71) |
| | | $C_3$ | |
| 1 | -0,96 | 67.2 | 0.95() - 0.11(69,69→75,75) |
| 2 | 1.62 | 11.8 | 0.91(70→71) - 0.15(70→72) + 0.13(70,70→71,72) + 0.11(67,70→71,73) |
| 6 | 3.04 | 0.7 | 0.49(70→72) - 0.47(67→71) - 0.45(68→71) + 0.25(70,70→71,71) - 0.19(67,70→71,71) |
| 8 | 3.42 | 2.3 | 0.54(70→72) + 0.44(64→71) + 0.38(67→71) + 0.31(70→73) - 0.16(70,70→71,72) |
| 9 | 3.51 | 1.0 | 0.51(64→71) + 0.47(70→74) - 0.36(70→73) - 0.21(70→76) - 0.17(70→75) + 0.15(68→72) |
| | | $C_{11}$ | |
| 1 | -1.01 | 48.9 | 0.91() - 0.19(70,70→71,71) - 0.17(70→71) - 0.11(69,69→76,76) |
| 2 | 0.54 | 34.1 | 0.81(70→71) + 0.36(70,70→71,71) + 0.25() - 0.12(70→72) - 0.10(70,70→72,72) |
| 3 | 1.43 | 1.7 | 0.45(70→72) + 0.40(68→71) - 0.39(70,70→71,72) - 0.31(70,70→71,71) + 0.20(69→71) + 0.19(70→71) |
| 6 | 1.88 | 2.0 | 0.73(68→71) - 0.36(67→71) + 0.27(68,70→71,71) - 0.19(70→72) - 0.15(70→71) + 0.13(70,70→71,72) |
| 7 | 2.23 | 2.6 | 0.47(66→71) + 0.43(67→71) - 0.40(70→72) + 0.28(66,70→71,71) - 0.22(70→75) + 0.13(68→71) |
| 8 | 2.70 | 2.2 | 0.43(70→75) + 0.39(70,70→71,72) + 0.37(66→71) - 0.36(70,70→71,75) + 0.22(67→71) + 0.20(70→74) |

| PTCDA | | | INDO-CI Calculation |
|---|---|---|---|
| # | Energy [eV] | Intensity [%] | Wavefunction |
| $O_{13}$ | | | |
| 1 | -1.19 | 87.6 | 0.94() - 0.11(69,69→77,77) - 0.10(66,66→74,74) |
| 2 | 1.80 | 4.9 | 0.92(70→71) + 0.12(66,70→71,74) + 0.12(67,70→71,71) |
| 10 | 4.36 | 0.8 | 0.79(68→72) - 0.26(66→72) + 0.13(67→71) - 0.13(70→73) + 0.12(61→72) + 0.11(66,68→72,74) |
| $O_{14}$ | | | |
| 1 | -0.99 | 69.8 | 0.94() - 0.11(69,69→76,76) |
| 2 | 1.07 | 17.4 | 0.86(70→71) + 0.28(70→72) + 0.18(70,70→71,72) + 0.14(70,70→71,71) |
| 6 | 2.51 | 0.8 | 0.80(68→71) - 0.22(70→72) + 0.18(70→71) + 0.18(68→72) + 0.16(66→71) + 0.15(68,70→71,71) |
| 7 | 2.75 | 2.6 | 0.58(66→71) + 0.40(70→72) + 0.29(67→71) + 0.28(70→75) + 0.19(70→73) - 0.18(70→71) |
| 12 | 3.63 | 1.4 | 0.41(70→72) - 0.35(70,70→71,71) - 0.21(70→75) - 0.19(70→77) - 0.19(66→72) - 0.15(66→71) |

**Table 2:**

| | $C_{11}$ | | INDO-CI Calculation |
|---|---|---|---|
| # | Energy-shift [eV] | relative Intensity [%] | Wavefunction |
| | | | **PTCDI** |
| 1 | 0.00 | 100.0 | 0.92() + 0.16(70,70→71,71) + 0.14(70→71) - 0.11(69,69→76,76) |
| 2 | 1.63 | 55.6 | 0.82(70→71) - 0.35(70,70→71,71) + 0.21() - 0.13(70→72) |
| 3 | 2.52 | 3.3 | 0.43(68→71) + 0.43(70→72) + 0.37(70,70→71,72) + 0.32(70,70→71,71) - 0.24(67→71) - 0.20(69→71) |
| 6 | 2.91 | 3.5 | 0.55(68→71) + 0.49(67→71) + 0.43(66→71) + 0.21(68,70→71,71) - 0.12(70→71) - 0.11(70→72) |
| 7 | 3.29 | 3.2 | 0.57(67→71) + 0.43(70→72) - 0.30(66→71) - 0.29(67,70→71,71) + 0.19(70→75) - 0.17(68→71) |
| 8 | 3.86 | 2.2 | 0.47(70→75) - 0.43(70,70→71,72) + 0.34(70,70→71,75) - 0.27(67→71) + 0.21(70→74) + 0.20(66→71) |
| | | | **PTCDA** |
| 1 | 0.00 | 100.0 | 0.91() - 0.19(70,70→71,71) - 0.17(70→71) - 0.11(69,69→76,76) |
| 2 | 1.55 | 69.7 | 0.81(70→71) + 0.36(70,70→71,71) + 0.25() - 0.12(70→72) - 0.10(70,70→72,72) |
| 3 | 2.44 | 3.4 | 0.45(70→72) + 0.40(68→71) - 0.39(70,70→71,72) - 0.31(70,70→71,71) + 0.20(69→71) + 0.19(70→71) |
| 6 | 2.89 | 4.0 | 0.73(68→71) - 0.36(67→71) + 0.27(68,70→71,71) - 0.19(70→72) - 0.15(70→71) + 0.13(70,70→71,72) |
| 7 | 3.24 | 5.4 | 0.47(66→71) + 0.43(67→71) - 0.40(70→72) + 0.28(66,70→71,71) - 0.22(70→75) + 0.13(68→71) |
| 8 | 3.71 | 4.6 | 0.43(70→75) + 0.39(70,70→71,72) + 0.37(66→71) - 0.36(70,70→71,75) + 0.22(67→71) + 0.20(70→74) |
| | | | **NDCA** |
| 1 | 0.00 | 100.0 | 0.93() - 0.12(36,36→37,37) |
| 2 | 1.93 | 22.6 | 0.80(36→37) - 0.34(35→37) - 0.26(36,36→37,37) + 0.12(35,36→37,37) - 0.12() + 0.10(36,36→37,38) |
| 3 | 2.03 | 3.3 | 0.81(35→37) + 0.32(36→37) - 0.23(35,36→37,37) - 0.15(34,35→37,37) + 0.13(29,35→37,37) - 0.12(36,36→37,37) |
| 4 | 3.47 | 8.6 | 0.84(34→37) - 0.23(34,36→37,37) - 0.18(34,34→37,37) - 0.18(35→38) |

## Table 3:

| | $O_{14}$ | | INDO-CI Calculation |
|---|---|---|---|
| # | Energy-shift [eV] | relative Intensity [%] | Wavefunction |
| | | | **PTCDI** |
| 1 | 0.00 | 100.0 | 0.95() - 0.11(69,69→76,76) |
| 2 | 2.18 | 19.5 | 0.90(70→71) - 0.14(70→72) + 0.15(70,70→71,71) - 0.14(70,70→71,72) - 0.11(67,70→71,74) + 0.11(66,70→71,72) |
| 6 | 3.57 | 0.9 | 0.81(68→71) - 0.24(70→72) + 0.21(66→71) - 0.16(67→71) + 0.15(68,70→71,71) - 0.12(70→71) |
| 7 | 3.74 | 1.4 | 0.64(66→71) + 0.47(70→72) + 0.26(70,70→71,72) - 0.26(70→73) + 0.21(70→75) + 0.15(70,70→71,71) |
| 12 | 4.67 | 1.9 | 0.33(66→72) + 0.31(70,70→71,71) + 0.30(70→72) - 0.24(66→71) + 0.23(70→73) - 0.22(70→78) |
| | | | **PTCDA** |
| 1 | 0.00 | 100.0 | 0.94() - 0.11(69,69→76,76) |
| 2 | 2.06 | 24.9 | 0.86(70→71) + 0.28(70→72) + 0.18(70,70→71,72) + 0.14(70,70→71,71) |
| 6 | 3.50 | 1.1 | 0.80(68→71) - 0.22(70→72) + 0.18(70→71) + 0.18(68→72) + 0.16(66→71) + 0.15(68,70→71,71) |
| 7 | 3.74 | 3.7 | 0.58(66→71) + 0.40(70→72) + 0.29(67→71) + 0.28(70→75) + 0.19(70→73) - 0.18(70→71) |
| 12 | 4.62 | 2.0 | 0.41(70→72) - 0.35(70,70→71,71) - 0.21(70→75) - 0.19(70→77) - 0.19(66→72) - 0.15(66→71) |
| | | | **NDCA** |
| 1 | 0.00 | 100.0 | 0.95() - 0.11(35,35→39,39) |
| 2 | 2.63 | 10.7 | 0.88(36→37) + 0.17(35→37) + 0.16(36,36→37,38) + 0.10(36,36→37,37) + 0.10(30,36→37,38) |
| 5 | 3.12 | 3.8 | 0.71(33→37) - 0.39(36→38) + 0.25(36→39) + 0.18(33,36→37,37) + 0.17(36→40) - 0.14(36,36→37,37) |
| 11 | 4.63 | 1.2 | 0.48(35→39) + 0.20(36→42) + 0.20(35→42) - 0.19(36→38) + 0.15(36→43) + 0.14(28→37) |

**Figure 1:**

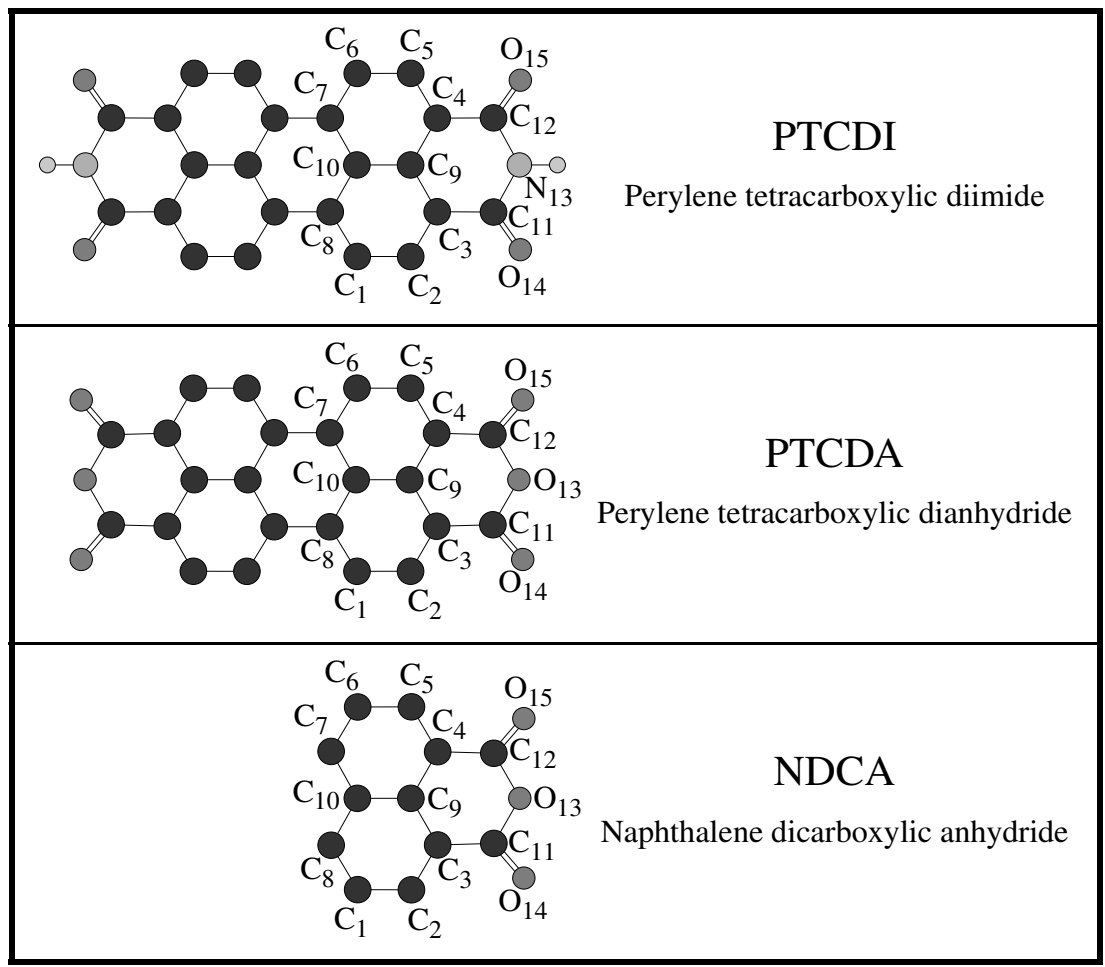

**Figure 2:**

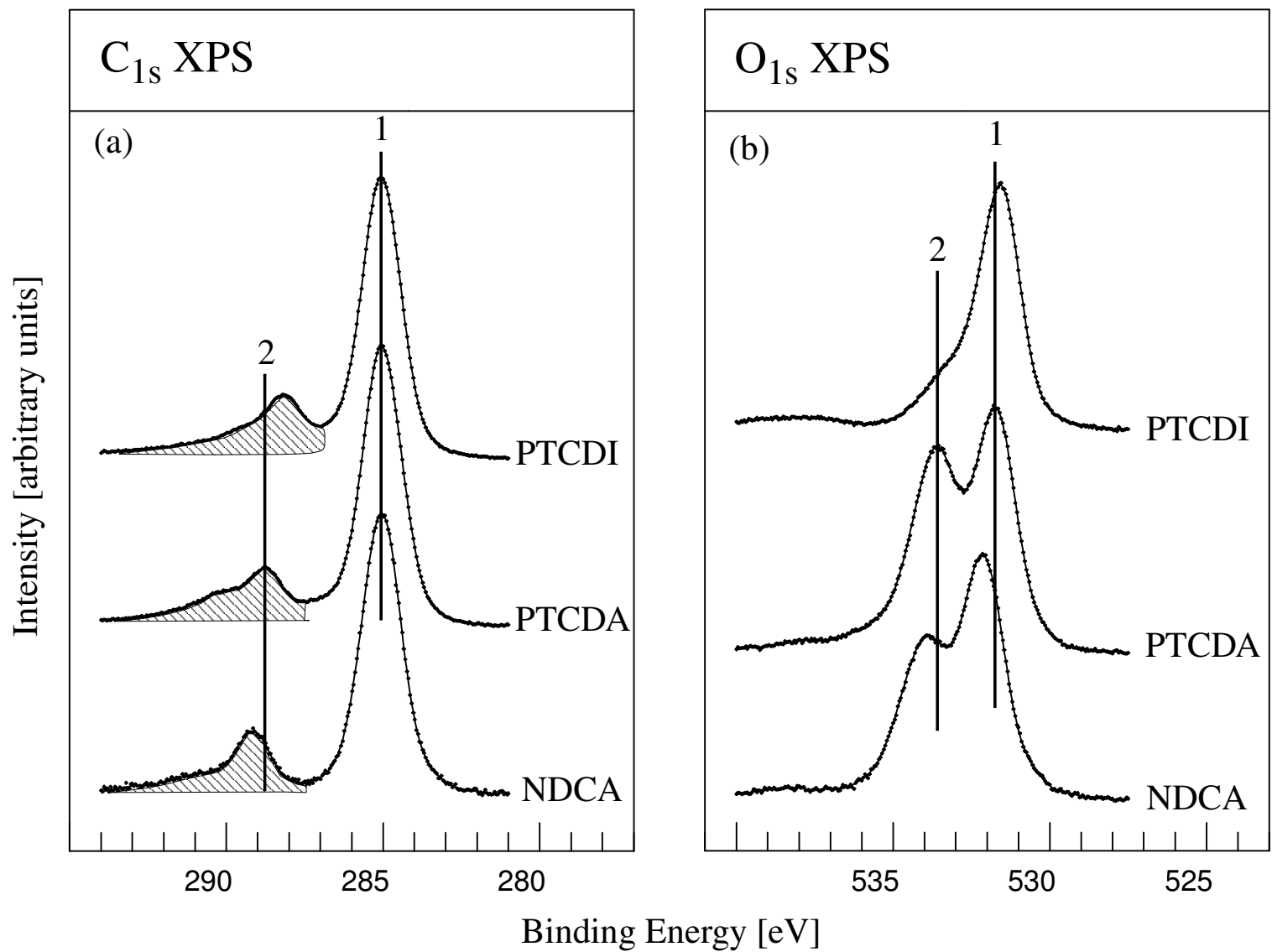

**Figure 3:**

| PTCDI | $D_{2h}$ | neutral | $C_s$ | $C_{11}$ ionized | $C_s$ | $O_{14}$ ionized |
|---|---|---|---|---|---|---|
| | 74 $B_{1u}$ ($\pi$) | | A" ($\pi$) | | A" ($\pi$) | |
| | 73 $B_{1u}$ ($\pi$) | | A" ($\pi$) | | A" ($\pi$) | |
| | 72 $A_u$ ($\pi$) | | A" ($\pi$) | | A" ($\pi$) | |
| LUMO | 71 $B_{3g}$ ($\pi$) | | A" ($\pi$) | | A" ($\pi$) | |
| HOMO | 70 $A_u$ ($\pi$) | | A" ($\pi$) | | A" ($\pi$) | |
| | 69 $B_{2g}$ ($\pi$) | | A" ($\pi$) | | A" ($\pi$) | |
| | 68 $B_{1u}$ ($\pi$) | | A" ($\pi$) | | A" ($\pi$) | |
| | 67 $B_{2g}$ ($\pi$) | | A" ($\pi$) | | A" ($\pi$) | |
| | 66 $B_{3g}$ ($\pi$) | | A" ($\pi$) | | A" ($\pi$) | |

**Figure 4:**

| PTCDA | $D_{2h}$ | neutral | $C_s$ | $C_{11}$ ionized | $C_s$ | $O_{14}$ ionized |
|---|---|---|---|---|---|---|
| | 74  $B_{1u}$ ($\pi$) | | A" ($\pi$) | | A" ($\pi$) | |
| | 73  $B_{1u}$ ($\pi$) | | A" ($\pi$) | | A" ($\pi$) | |
| | 72  $A_u$ ($\pi$) | | A" ($\pi$) | | A" ($\pi$) | |
| LUMO | 71  $B_{3g}$ ($\pi$) | | A" ($\pi$) | | A" ($\pi$) | |
| HOMO | 70  $A_u$ ($\pi$) | | A" ($\pi$) | | A" ($\pi$) | |
| | 69  $B_{2g}$ ($\pi$) | | A" ($\pi$) | | A" ($\pi$) | |
| | 68  $B_{1u}$ ($\pi$) | | A" ($\pi$) | | A" ($\pi$) | |
| | 67  $B_{2g}$ ($\pi$) | | A" ($\pi$) | | A" ($\pi$) | |
| | 66  $B_{3g}$ ($\pi$) | | A" ($\pi$) | | A" ($\pi$) | |

**Figure 5:**

| NDCA | | $C_{2v}$ | neutral | $C_s$ | $C_{11}$ ionized | $C_s$ | $O_{14}$ ionized |
|---|---|---|---|---|---|---|---|
| | 40 | $A_2$ ($\pi$) | | A" ($\pi$) | | A" ($\pi$) | |
| | 39 | $B_2$ ($\pi$) | | A" ($\pi$) | | A" ($\pi$) | |
| | 38 | $B_2$ ($\pi$) | | A" ($\pi$) | | A" ($\pi$) | |
| LUMO | 37 | $A_2$ ($\pi$) | | A" ($\pi$) | | A" ($\pi$) | |
| HOMO | 36 | $A_2$ ($\pi$) | | A" ($\pi$) | | A" ($\pi$) | |
| | 35 | $B_2$ ($\pi$) | | A" ($\pi$) | | A" ($\pi$) | |
| | 34 | $B_2$ ($\pi$) | | A" ($\pi$) | | A' ($\sigma$) | |
| | 33 | $A_1$ ($\sigma$) | | A' ($\sigma$) | | A" ($\pi$) | |
| | 32 | $B_1$ ($\sigma$) | | A' ($\sigma$) | | A' ($\sigma$) | |

**Figure 6:**

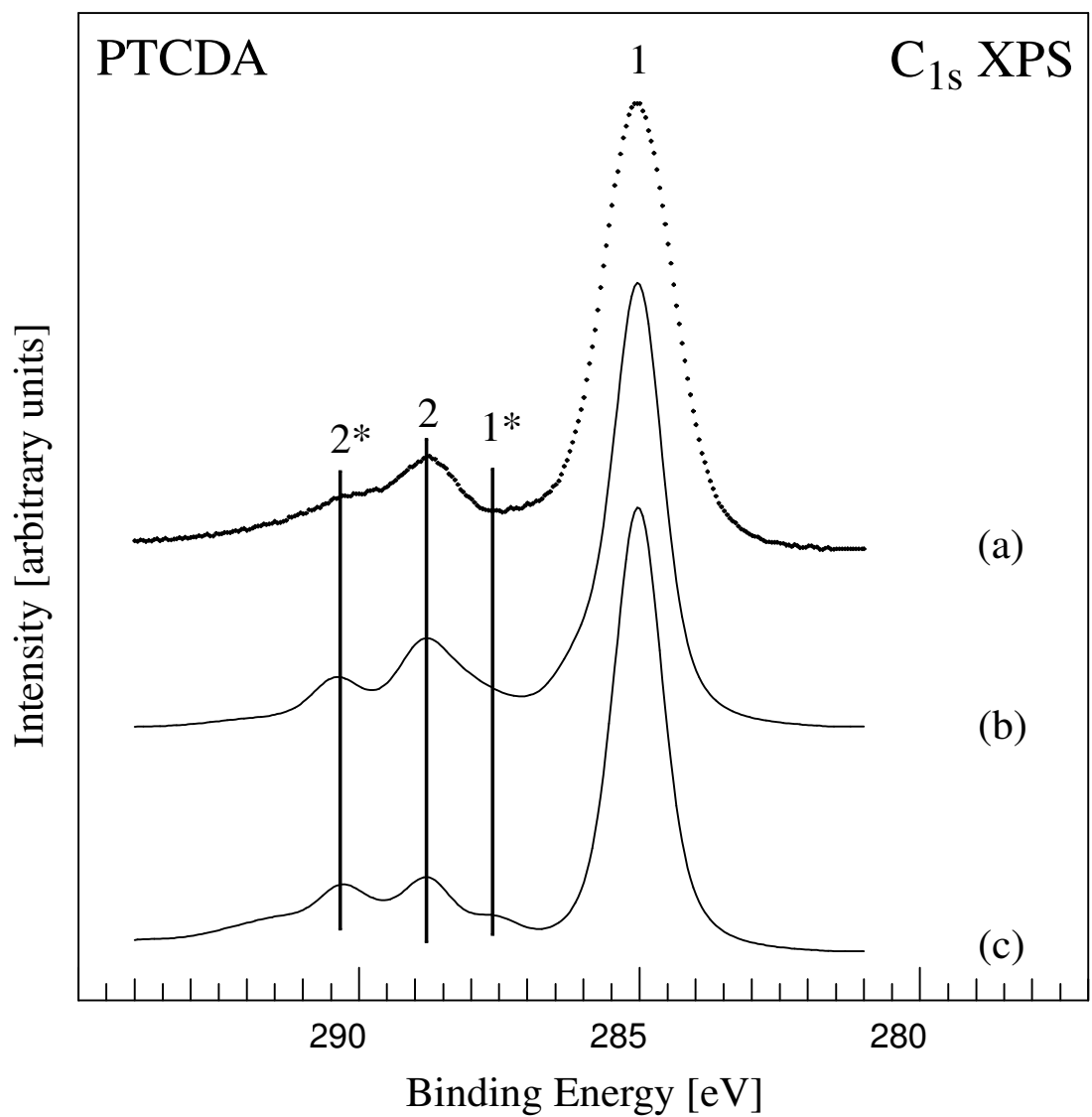

**Figure 7:**

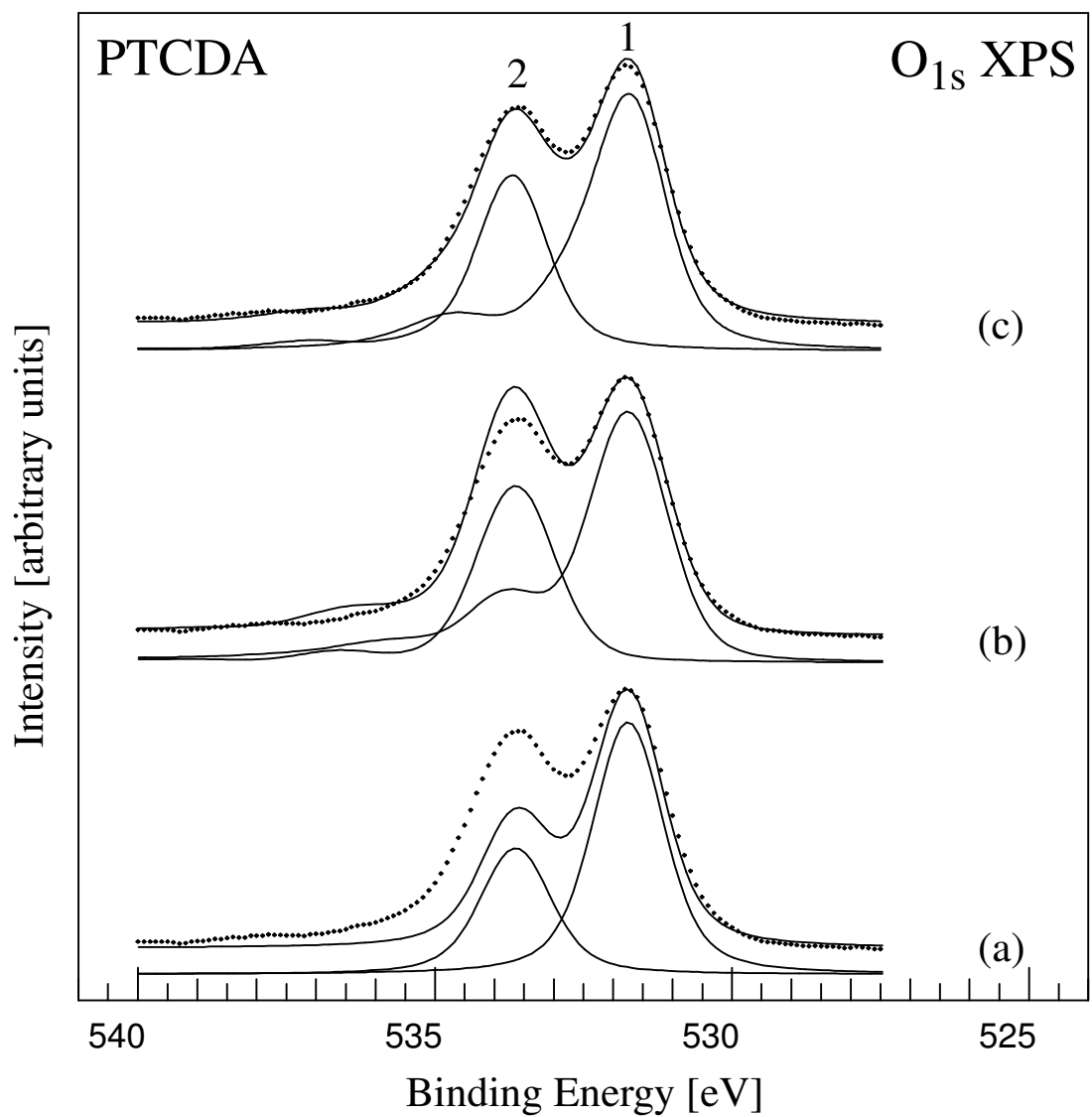